\documentclass{scrartcl}
\usepackage{amsmath}
\usepackage{amssymb}
\usepackage{graphicx}
\usepackage{psfrag}
\usepackage[dvips]{color}
\usepackage{MnSymbol}

\begin{document}

\title{Would be the photon a composed particle? quantization of field fluxes in electromagnetic radiation}

\subtitle{English translation of the Portuguese version ``Sobre a possibilidade da quantiza\c c\~ao dos fluxos de campo na radia\c c\~ao eletromagn\'etica'', published in Rev. Bras. Ens. F\'is., \textbf{35} (2013) p. 3305}

\author{Celso de Araujo Duarte\footnote{celso@fisica.ufpr.br}\\ Departamento de F\'isica, Universidade Federal do Paran\'a\\ CP 19044, 81531-990, Curitiba, PR, Brazil}

\date{}
%keywords: flux quantization; photon; electromagnetic radiation; magnetic flux; magnetic monopole; electric flux

\maketitle
\begin{abstract}
Here it is made a comparative analysis between the classical and the quantum expressions for the energy of electromagnetic radiation (ER). The comparison points to the possibility of the quantization of the magnetic and the electric field fluxes in the ER.
\end{abstract}

\section{Introduction}

It is worth noting that the seeds of quantum mechanics (QM) lie on the 1901 Planck's work on the blackbody radiation, which for the first time introduced the concept of quantization of energy. This concept provided an unexpected dependence of the quanta of energy on the frequency of the radiation \cite{planck}, which entered complete disagreement with the conceptions arisen from the classical electrodynamics, since according to that the intensity of radiation $I$ depends on the square of the amplitude of the electric field ($\boldsymbol E$) \cite{jackson}.

Being the probabilistic character the most striking peculiarity of this new science (QM) that governs the microscopic world, but the origin of its name stems on just the concept of quantization. After Planck's work, Einstein sustained the Planck's energy quantization hypothesis on his 1905 work concerning on the physical explanation of the photoelectric effect \cite{einstein}, with the introduction of the concept of the photon -- the quantum of light. This new physical entity was shown by Einstein to be essential to the comprehension of the sensitivity of materials to the light irradiation as verified on experiments involving the measurement of photo currents. The framework of physics based on the concepts of the classical electrodynamics was completely unsuccessful on the attempts to give an explanation to that experimental findings.

Presently, the concept of the photon remains as on its birth, as a testimony of another impressive outcome of QM: the particle-wave duality. The historical evolution brought new features, as the spin that is associated to the wave helicity and angular momentum, and the linear momentum inherited from the classical electromagnetic wave counterpart.

The above-mentioned dependence of the photon energy on its associated frequency remains an impressive fact. However, an integration between the conflicting classical and quantum pictures of the electromagnetic radiation (ER) was achieved by the understanding that the macroscopic classical, wavelike radiation -- whose intensity (the energy flux density) is proportional to $\boldsymbol{E}^2$ --, can be considered as a flux of photons per unit area per unit time, where each photon carries an amount of energy $h\nu$ proportional to its intrinsic frequency \cite{einstein,messi}.

It can then be considered that this question was overcome, and the wave-particle duality was satisfactory understood with respect to the classical versus quantum picture conflict mentioned on the previous paragraph. However, we can explore this question further trying to improve the comprehension of the integration between the particle and the wave aspects. In the present work, it is presented a study of the classical \textit{versus} the quantum pictures of light, which suggests the possibility that the photon itself can be considered as a state comprised by two distinct particles, a quantum of electric field flux (qEF) and a quantum of magnetic field flux (qMF).

This work is organized as follows: in section \ref{initialconsid} it is shown the equivalence  between the classical energy flux density and the quantum photon energy, which opens the possibility to the introduction of the concepts of the electric and the magnetic field flux quanta. In section \ref{quanta}, it is proposed that these flux quanta qEF and qMF have an inherent inertia and a moment of inertia. Section \ref{fluctuations} uses the uncertainty relations involving the components of the electric and the magnetic fields, which point to the existence of a magnetic charge. As a corollary, it is shown an estimation of the energy fluctuations of the ER in the presence of a charge density. Section \ref{bound} continues with uncertainty relations and provides a lower bound for the value of the product of the mean electric and magnetic field flux quanta. Finally, sections \ref{discus} and \ref{concl} present the discussions and the final conclusions.

\section{Initial considerations}\label{initialconsid}

We start on the grounds of the classical electrodynamics, with the modulus of the Poynting vector $\boldsymbol{S}$ which expresses the energy flux density per unit area per unit time \cite{jackson},
\begin{equation}\label{poynt}
S=\frac1{\mu_0}\left|\boldsymbol{E}\times\boldsymbol{B}\right|,
\end{equation}
where $\boldsymbol{E}$ and $\boldsymbol{B}$ are respectively the electric and the magnetic induction field vectors (on the SI units).

\begin{figure}[ht]
\begin{center}
\includegraphics[width=.5\textwidth]{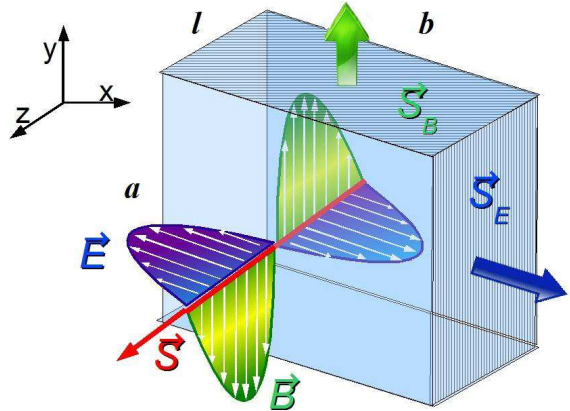}
\end{center}
\caption{(color online) Schematic representation of the electric field $\boldsymbol{E}$ and the magnetic field $\boldsymbol{B}$ of a plane sinusoidal wave of ER. The rectangular box encloses a half wavelength, and has a surface of sides \textit{a} and \textit{b} perpendicular to direction \textit{z} of the wave propagation (the direction of the Poynting vector $\boldsymbol{S}$) and surfaces of sides \textit{a}, \textit{l} and area $S_E=al$ perpendicular to the electric field $\boldsymbol{E}$ and sides \textit{b}, \textit{l} and area $S_B=bl$ perpendicular to the magnetic field $\boldsymbol{B}$.}
\label{fig1}
\end{figure}

Let us consider a sinusoidal plane wave. In the geometry presented on figure \ref{fig1}, we represent a half-wavelength inserted on a rectangular box with side lengths $a$, $b$ and $l$, whose face of surface area $ab$ is perpendicular to $\boldsymbol{S}$ and the faces of surface areas $S_E=al$ and $S_B=bl$ are perpendicular to $\boldsymbol{E}$ and $\boldsymbol{B}$, respectively. Now we write this expression in terms of the mean electric and magnetic field fluxes $\left\langle \Phi_E\right\rangle$, $\left\langle \Phi_B\right\rangle$ flowing throughout the corresponding perpendicular cross sectional areas $S_E$ and $S_B$ (note that $\boldsymbol{E}$ and $\boldsymbol{B}$ are mutually perpendicular),
\begin{equation}\label{poynt3}
\left\langle S\right\rangle=\frac{\left\langle \Phi_E\Phi_B\right\rangle}{\mu_0S_ES_B}=\frac{\left\langle \Phi_E\right\rangle\left\langle \Phi_B\right\rangle}{\mu_0l^2ab}
\end{equation}

Because $\left\langle S\right\rangle ab$ is the mean energy flux per unit time $W$, then $W=\left\langle \Phi_E\right\rangle\left\langle \Phi_B\right\rangle/(\mu_0l^2)$. However, $l=\lambda/2= c/2\nu$, where $c$ is the velocity of light on vacuum and $\nu$ is the frequency. As a consequence,  the mean flux energy can be written as $W=4\left\langle \Phi_E\right\rangle\left\langle \Phi_B\right\rangle/(\mu_0 c^2)\nu^2$. Being $W/2\nu=E_{\lambda/2}$ the overall energy carried by a half-wavelength of ER during a half period $T=1/2\nu$, we have
\begin{equation}\label{poynt6}
E_{\lambda/2}=\frac{2\left\langle \Phi_E\right\rangle\left\langle \Phi_B\right\rangle}{\mu_0c^2}\nu
\end{equation}
Note that this expression is proportional to the frequency. This leads us to compare directly to the quantum counterpart, where the photon energy is given by the well-known expression \cite{einstein,messi}
\begin{equation}\label{hnu}
E=h\nu
\end{equation}
Then, the equality between equations \ref{poynt6} and \ref{hnu} requires that $2\left\langle \Phi_E\right\rangle\left\langle \Phi_B\right\rangle/\mu_0c^2=h$. We can consider the possibility that $\left\langle \Phi_B\right\rangle$ can be identified with a quantum of magnetic field flux, qMF, and also identify $\left\langle \Phi_E\right\rangle$ with a quantum of electric field flux, qEF. Since from the integral form of the Gauss law \cite{jackson} the electric field flux of a point charge \textit{q} is simply $q/\epsilon_0$, we can suppose that the electric field flux is quantized in terms of the unit charge \textit{e}, so qEF could be $\phi_E=e/\epsilon_0$. Consequently the supposition $\left\langle \Phi_E\right\rangle=\phi_E$ implies directly in
\begin{equation}\label{monop}
\left\langle \Phi_B\right\rangle=\frac h{2e}
\end{equation}
which can be considered similarly the qMF $\phi_B$. Note that the magnitudes of these field quanta are related by the fine-structure constant $\alpha$ -- apart from the dimensionality factor being the velocity of light $c$ --, or in other words, $\phi_E=4c\alpha\phi_B$. The quantity $\phi_B$ is exactly equal to the field flux arisen from a Dirac magnetic monopole (MP) \cite{dirac}.

\section{The quanta of electric and magnetic field fluxes in ER}\label{quanta}

The previous section served as a starting point for more detailed speculation. First, note that the values of $\left\langle \Phi_E\right\rangle$ and $\left\langle \Phi_B\right\rangle$ were intentionally calculated for a single half wavelength, but one could argue that in a similar manner, any multiple value of this magnitude could be chosen. Actually, a better acceptable choice could be the overall extent of the single wavepacket of a photon; considering this as, for example, equal to the pulse duration of a particular bandwidth (for example, of an electronic transition on an atom) multiplied by the velocity of light. Certainly, this extent would not be equal to an integer multiple of a single half wavelength, but we can consider by hypothesis that the overall field fluxes $\left\langle \Phi_E\right\rangle$ and $\left\langle \Phi_B\right\rangle$ along the entire extent would be such that the product $\left\langle \Phi_E\right\rangle\left\langle \Phi_B\right\rangle$ is in accordance to the equality between equations \ref{poynt6} and \ref{hnu}. In this sense, while the classical theory considers the ER as a purely wavelike phenomenon, the quantum nature of light could be considered of quanta of energy (the photons) carried by pairs of quanta of field fluxes.

In electrostatics and magnetostatics, it is a well-known fact that the electric and the magnetic fields store energy in the space, where the energy density per unit volume is proportional to the square of amplitude of $\boldsymbol E$. As a consequence, it is not surprising that the ER carries energy. On the counterpart, we can express this reverse, that since the ER carries energy the fields also store energy.

The same cannot be said about linear momentum. The ER carries momentum; however there is not momentum neither in static electric nor in magnetic fields. This difficulty can be overcome if we consider that the fields and field fluxes have an inertia, so that despite static fields do not carry moment the ER does. Then, the conception that the light carries inertia would be conceptually related to well known idea that the inertia of matter is dependent on its energy content \cite{einstein2}. In this way similar arguments can be said about the zero and nonzero angular momentum of static fields and the ER, respectively, say that the field flux quanta, and consequently also does the electromagnetic field, have an intrinsic moment of inertia. 

Note that, despite for an electromagnetic wave the classical electrodynamics requires that the electric and the magnetic fields are related by $E=cB$, a similar relation involving the field fluxes cannot be established, since the areas $S_E$ and $S_B$ are not determined.

\section{Energy fluctuations in the ER in the presence of a charge density}\label{fluctuations}

Let us consider that $\psi$ is the wavefunction of the photon, which as well known satisfies the wave equation $(\nabla^2-c^{-2}\partial^2_t)\psi=0$. We can apply the chain rule \cite{stewart} to verify the following identity involving the components of the linear momentum operator $\hat p$ \cite{messi,landau} and the electric field operator $\vec{\hat E}$:
\begin{equation}\label{uncertE}
\hat{p_i}\hat{E_i}-\hat{E_i}\hat{p_i}=i\hbar\partial_i E_i=\left[\hat{p_i},\hat{E_i}\right]
\end{equation}
Being the commutator nonzero, we have the following uncertainty relations involving the values of the corresponding components of linear momenta and electric field:
\begin{equation}\label{uncertE2}
\delta p_i\delta E_i\geq \hbar\alpha_i
\end{equation}
where the $\alpha_i$ are given by
\begin{equation}\label{alphai}
\alpha_i=\int\psi^*\partial_i E_i\psi d^3x
\end{equation}
However, from the Gauss law $\sum\partial_i E_i=\rho/\epsilon_0$ ($\rho$ is the electric charge density). As a consequence, 
\begin{equation}\label{uncertE3}
\sum_{i=1}^3\delta p_i\delta E_i\geq \hbar\bar{\rho}
\end{equation}
where we have written $\bar{\rho}=\int\psi^*\rho\psi d^3x$. A similar relation can be written for the magnetic field,
\begin{equation}\label{uncertB}
\sum_{i=1}^3\delta p_i\delta B_i\geq 0
\end{equation}
For the electromagnetic radiation, $E=cB$, and as a consequence the uncertainty relation in magnetic field \ref{uncertB} cannot satisfy the uncertainty relation in electric field \ref{uncertE3}. In other words, the lower bounds of the field fluctuations are compatible only if $\sum_{i=1}^3\delta p_i\delta B_i\geq c\hbar\bar{\rho}$. Since the nonzero value of the right member of relation \ref{uncertE3} arises from the nonzero charge density in Gauss law, we are led to believe on the necessity of a magnetic charge density such that we have
\begin{equation}\label{uncertB2}
\sum_{i=1}^3\delta p_i\delta B_i\geq \frac{\hbar \bar{\rho}}c
\end{equation}
(note that the fields are the total electric $\vec E$ and magnetic $\vec B$ fields, say that the sum of the fields of the ER and the fields generated by the charge distribution). The rough estimation of the fluctuations of the energy flux is $\delta S\sim \delta E\delta B/\mu_0$ considering $\vec E$ and $\vec B$ about perpendicular (here we do not consider the fields arisen from the charge density), so in this sense the product of equations \ref{uncertE3} and \ref{uncertB2} give (approximately)
\begin{equation}\label{uncertPoynt}
\delta p^2\delta S\geq \frac{\left(\hbar \bar{\rho}\right)^2}{\mu_0c}
\end{equation}
which gives a lower bound for the fluctuations of the flux of energy that does not depend directly on the frequency of the ER, as would be expected considering the vacuum energy of the quantum harmonic oscillator \cite{messi,loudon}.

\section{Lower bound for the flux quanta}\label{bound}

We can make a consideration about the value the product $\left\langle \Phi_E\right\rangle\left\langle \Phi_B\right\rangle$ (present in equations \ref{poynt3} and \ref{poynt6}) employing uncertainty relations. The uncertainty relation for the components of the electric and the magnetic field of the ER are given by \cite{heisen} (here in SI units)
\begin{equation}\label{uncert1}
\delta E\delta B \geq \mu_0\frac{hc^2}{vl}
\end{equation}
where $v$ is the considered volume and in our case, $v=abl=ab\lambda$. Replacing $\delta E$, $\delta B$ by $\left\langle \Phi_E\right\rangle/S_E$, $\left\langle \Phi_B\right\rangle/S_B$,
\begin{equation}\label{uncert2}
\frac{\left\langle \Phi_E\right\rangle\left\langle \Phi_B\right\rangle}{S_ES_B} \geq \mu_0\frac{hc^2}{vl}
\end{equation}
or
\begin{equation}\label{uncert3}
\frac{\left\langle \Phi_E\right\rangle\left\langle \Phi_B\right\rangle}{\mu_0c^2}\geq h
\end{equation}
which express that there is a lower bound for the product $\left\langle \Phi_E\right\rangle\left\langle \Phi_B\right\rangle$, determined only by universal constants. The inequality \ref{uncert3} leaves us immediately to the comparison between equations \ref{poynt6} and \ref{hnu}. Note that the choices $\left\langle \Phi_E\right\rangle=e/\epsilon_0$ and $\left\langle \Phi_B\right\rangle=h/e$ lead to the equality sign on the above equation \ref{uncert3}.

\section{Discussions}\label{discus}

Here we introduce comments about the exposed above. In section \ref{fluctuations} it was pointed that the equivalence between the electric field uncertainty relation \ref{uncertE3} requires equivalently the magnetic field uncertainty relation \ref{uncertB2}. Following the reverse way of the obtain of the electromagnetic field uncertainty relation \ref{uncertE3}, we infer that the divergence of the magnetic field $\boldsymbol{B}$ is nonzero, or in other words, there is a MP charge as concluded on that section. This question is somewhat out of the scope of the present work, where instead the magnetic field flux is relevant. However, the section \ref{fluctuations} reinforces the hypothesis of the existence of the MP. Both the discussions of the last section (\ref{bound}) and the initial considerations at section \ref{initialconsid} are suggestively in accordance with the value of the Dirac's MP of charge $\mu_0h/(2e)$ \cite{dirac} which corresponds to the magnetic field flux $\phi_0=h/(2e)$, when one considers that the electric field flux is also quantized in units of $e/\epsilon_0$, which is the flux of the unit charge $e$.

It is worth noting that the qMF is a well-known physical magnitude. The concept of quantization of the magnetic field flux appears within the context of solid-state physics within the theories of superconductivity (SC) \cite{tinkham} and the fractional quantum Hall effect (FQH) \cite{jain} (the quantization of magnetic field flux was discovered by Deaver and Fairbank \cite{deav-fair} and Doll and Nabauer \cite{doll-nab}. In the framework of the SC the quantum of magnetic flux is $\phi_{SC}=h/(2e)$, and on the composite fermions theory for the FQH \cite{jain} we have $\phi_{FQH}=h/e$ , which differs from the previous by the factor of two). While the qMF and the electric unit charge are well-know physical entities, it is fair to consider the existence of their counterparts -- the unit magnetic charge and the quantum of electric field flux.

These arguments together with the results shown in section \ref{initialconsid} allow us to consider the possibility that the photon itself is composed by the elementary quanta of electric and magnetic field fluxes.

\section{Conclusions}\label{concl}

As a conclusion, it was shown a mathematical relation between the classical and the quantum expressions for the energy of the ER. The comparison points to the existence of the quanta of electric and magnetic field fluxes as constituent particles of the photon that presumably have intrinsic inertia.

\end{document}